**Title:** Production of Martian fiber by in-situ resource utilization strategy

**Authors:** Ze-Shi Guo[1, 2], Dan Xing[1, 2], Xiong-Yu Xi[1, 2], Cun-Guang Liang[1, 2], Bin Hao[1, 2], Xiaojia Zeng[3], Hong Tang[3], Huaican Chen[4, 5], Wen Yin[4, 5], Peng Zhang[6], Kefa Zhou[6], Qingbin Zheng[7], Peng-Cheng Ma[1, 2, *]

**Affiliations:**
1 Laboratory of Environmental Science and Technology, The Xinjiang Technical Institute of Physics and Chemistry, Key Laboratory of Functional Materials and Devices for Special Environments, Chinese Academy of Sciences, Urumqi, 830011, China.
2 Center of Materials Science and Optoelectronics Engineering, University of Chinese Academy of Sciences, Beijing, 100049, China.
3 Center for Lunar and Planetary Sciences, Institute of Geochemistry, Chinese Academy of Sciences, Guiyang, 550081, China.
4 Institute of High Energy Physics, Chinese Academy of Sciences, Dongguan, 523000, China.
5 Spallation Neutron Source Science Center, Dongguan, 523803, China.
6 Technology and Engineering Center for Space Utilization, Chinese Academy of Sciences, Beijing, 100094, China.
7 School of Science and Engineering, The Chinese University of Hong Kong, Shenzhen, Shenzhen, 518172, China.
*Corresponding author. Email: mapc@ms.xjb.ac.cn

**Abstract:** Many countries and commercial organizations have shown great interest in constructing Martian base. In-situ resource utilization (ISRU) provides a cost-effective way to achieve this ambitious goal. In this paper, we proposed to use Martian soil simulant to produce fiber to satisfy material requirement for the construction of Martian base. The composition, melting behavior and fiber forming process of soil simulant was studied, and continuous fiber with a maximum strength of 1320 MPa was obtained on a spinning facility. The findings of this study demonstrate the feasibility of ISRU to prepare Martian fiber from the soil on the Mars, offering a new way to get key materials for the construction of Martian base.

**Introduction**

Mars is the fourth planet from the Sun and exhibits significant similarities to the Earth. It is considered to be the most suitable site for human migration in the solar system, as well as a transit station for deep space exploration [1, 2]. Since the cold war, the US and former Soviet Union launched probes to Mars and put forward the idea of a Martian base with some prospective studies [1-4]. More recently, NASA's Perseverance rover landed on Martian surface and successfully flown a helicopter for survey [5]. NASA plans to start the first human mission to Mars after finishing the Artemis program, which is considered to be the preparation for the human Mars exploration, aiming at constructing a lunar camp that can be used to launch and supply spacecraft to Mars [6]. China launched its first Mars probe, Tianwen-1, with an orbiter, a lander, and a rover in 2020. The rover Zhurong has landed on Mars and started a serial of new exploration activities [7, 8]. China also set up a base to simulate the environment of Mars for future research [9]. While there is no exact timetable for the construction of Martian base proposed by the leading countries, they all showed great interest in achieving this giant ambition. Some commercial organizations also made quite aggressive plan for the Martian base construction. For example, SpaceX proposed to build a city and send 1 million people to Mars by 2050. Blue Origin also expressed an interest in building a base on the Mars [10]. From the available information, it seems that the construction of a Martian base is only a matter of time. For the construction of the Martian base, it requires huge amount of building materials. Considering the high cost and long distance from the Earth to the Mars, it is impossible to transport all these materials from the former. So how to develop a manufacturing system on Mars is one of the most challenging disciplines in both scientific and engineering fields [11], and in-situ resource utilization (ISRU) is an effective way [12, 13]. Mars surface is covered with a layer of soil, and the abundant amount of such soil makes it ideal for ISRU. As Martian soil is not available so far on the Earth, corresponding simulant is used as a substitute for ISRU studies. Previous studies for Martian ISRU often focused on materials with bulky structures, like concrete or metal alloy [1, 14-16]. With considerations on the harsh environment on the Mars, those traditional materials may not be versatile enough to satisfy the structural and functional requirements for base construction. In this context, developing novel materials with tailored performance is highly desirable. One of the ideal materials is composites, like the steel-reinforced concrete, a common construction material on the Earth. Composites are a binary system consisting of matrix and reinforcement [17]. Usually, the reinforcement has better mechanical properties and plays a key role in governing the performance of composites, whereas the matrix in the composites is responsible for bonding the reinforcement and protecting it from the environment.

Basalt fiber is a kind of inorganic filament made from basalt rock. This fiber has attracted great attention as reinforcement in construction material on the Earth [18, 19]. Martian soil has a similar composition and mineralogy to the basalt on the Earth [20], so if we could use such soil to produce corresponding fiber, and the obtained material had the potential to be used as reinforcement for the construction on the Mars. In other words, combination of Martian fiber with corresponding soil provides a new way to address the material challenges for the construction of Mars base.

While some groups and organizations have put forward the concept of Mars base, marginal progress has been made so far focusing on offering material solution to achieve this. The purpose of this paper is to demonstrate the feasibility of producing Martian fiber (MF) and evaluate its performance.

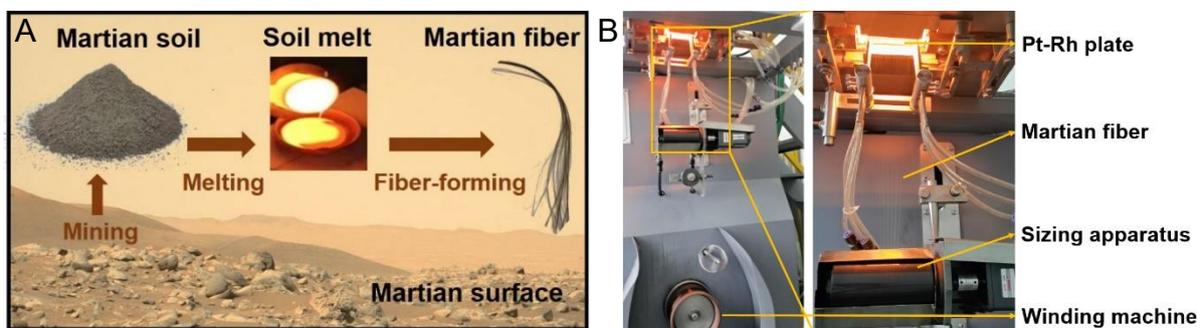

**Fig. 1.** Experimental setups for the production of Martian fiber (A: Technical design of a two-step procedure for fiber spinning, and the pictures are the states of material at different experimental setups; B: Facility for the production of continuous fiber).

**Materials and Methods**

**Materials and chemicals**

Martian regolith simulant (JMSS-1) was used in this study, and it was supplied by the Institute of Geochemistry, Chinese Academy of Sciences. Ultrapure water was prepared in the laboratory and all other reagents used in this research were in analytical grade.

**Characterization**

The morphology of materials was characterized by a laser scanner microscope (VK-X1000, Keyence, Japan) and scanning electron microscopy (SEM, Phenom XL, Phenom Scientific, US). The sample for SEM characterization was gold-plated to enhance the conductivity. The major element composition of material was analysed by X-ray fluorescence (XRF, S8 TIGER, Bruker, Germany). The crystal and mineral phases in the material were characterized by X-ray diffraction (XRD, D8, Bruker, Germany). The change of mass and heat flow of Martian simulant was characterized by differential scanning calorimetry (DSC, STA449F3, Netzch, Germany). The viscosity of melt was obtained on a high-temperature rotary viscometer (DV-III, Brookfield, USA). The tensile strength of single fiber was tested on a fiber tensile tester (XQ-1A, Shanghai Xinxian, China). The tests were performed according to the ASTM C1557-14 standard, the gauge length and loading speed were fixed at 25 mm and 2 mm/min, respectively. The microstructure of fiber was characterized by Raman spectroscopy (Lab RAM HR Evolution, HORIBA Scientific, France) with a green laser (532 nm). The functional groups on fiber surface were characterized by the Fourier transform infrared spectroscopy (FT-IR, Nicolet iS50, Thermo Fisher, USA). Neutron diffraction experiments were performed on a Multi-physics Instrument established in China Spallation Neutron Source, which is a total scattering neutron time-of-flight diffractometer [21]. Structure refinement was performed using the General Structure Analysis System (GSAS) II program based on the Rietveld method [22].

**Results and discussion**

A two-step procedure was established for the production of MF (Fig. 1A). In a typical process, Martian simulant (250.0 g) was put into a corundum crucible and melted in an electrical furnace to homogenize the starting material. The temperature was raised from room temperature to 1200 °C at a heating rate of 5 °C/min and then increased to 1500 °C at a heating rate of 2 °C/min. The melt was quenched by water to get the material in a glass state. Then the glass was crushed and added into the furnace of a fiber spinning facility established in our lab (Fig. 1B). During the experiment, the melted glass at around 1250 °C was continuously drawn from the platinum-rhodium alloy bushing with 50 nozzles, and water was used as a sizing agent to facilitate the winding of the filament. Continuous fibers with different diameters were obtained by controlling the speed of the winding machine with 2.5, 4.0 and 6.0 m/s, and the obtained fiber was dried in an oven at 120 °C for 12 h.

The initial study was focusing on getting information on the properties of Martian soil simulant. Images from the scanning electronic microscope (SEM) confirmed that the grain size of this simulant is below 1 mm (Fig. 2A), and the mean particle size is about 250 μm. The typical particle in the sample showed a rough and uneven granular surface with many edges and corners (B and C in Fig. 2), which is the result of a mechanical comminution process. This process was close to the physical weathering on Mars, where wind abrasion and meteoric impact led to the formation of the Martian soil [20]. The simulant has a similar composition to the Martian basaltic soil (Table 1). In both samples, $SiO_2$ is the major composition with around

50% of weight, followed by $Al_2O_3$, $Fe_xO_y$, MgO and CaO. In fiber science and technology, the acidity modulus ($M_k$), as defined by the $(W_{SiO2}+W_{Al2O3})/(W_{MgO}+W_{CaO})$, is used to evaluate the possibility of material to form filament, and the optimal $M_k$ for continuous basalt fiber is in the range of 3.0-6.0 [23]. The calculated $M_k$ for the average Martian soil and simulant are 3.7 and 4.1, respectively, suggesting the suitability of Martian soil to prepare fiber.

**Table 1.** Comparison on the chemical composition of Martian soil and the simulant [20].

| Composition (wt%) | $SiO_2$ | $Al_2O_3$ | $Fe_xO_y$ | MgO | CaO | $Na_2O$ | $K_2O$ | $TiO_2$ | MnO | $P_2O_5$ |
|---|---|---|---|---|---|---|---|---|---|---|
| Soil average | 45.41 | 9.71 | 16.73 | 8.35 | 6.37 | 2.73 | 0.44 | 0.91 | 0.33 | 0.83 |
| Simulant | 49.83 | 13.83 | 15.41 | 6.79 | 8.68 | 2.76 | 0.99 | 1.70 | 0.01 | 0.01 |

The crystal phase in the simulant is illustrated by an X-ray diffractometer (XRD), and it is found that the material is a mixture of several minerals (Fig. 2D), such as augite (Ca(Mg, Fe, Al)[(Si, Al)$_2O_6$], PDF#01-073-8541), ilmenite ($FeTiO_3$, PDF#04-006-6575), forsterite ($Mg_2SiO_4$, PDF#97-006-4739), labradorite ((Na,Ca)(Al,Si)$_4O_8$, PDF#04-011-6816), magnetite ($Fe_3O_4$, PDF#04-006-6550) and hematite ($Fe_2O_3$, PDF#04-003-2900). These results were in good agreement with those obtained from launched probes in real Martian soil [20]. The XRD result of the quenched sample shows a broad hump in the $2\theta$ range of 20°-30° (Fig. 2D), indicating the transformation of crystal structures in the simulant to the amorphous state.

Differential scanning calorimetry (DSC) technique gives information on the thermal behavior of soil under elevated temperature. The first endothermic peak at approximately 100 °C is attributed to the volatilization of water adsorbed in the sample, and the last one at around 1360 °C means the complete melting of the sample under the elevated temperature (Fig. 2E). Interestingly, several exothermic peaks are also noticed in the sample, suggesting the crystallization or oxidation of crystals in the melt. For example, the one at 1185 °C is attributed to the formation of the hematite in the melt [24]. Above this temperature, there will be absence of crystals in the material, suggesting the formation of entirely glassy state of material. The viscosity of the simulant was significantly decreased from 1500 to 250 dPa·s within a narrow temperature range of 1210-1260 °C (Fig. 2F). It kept nearly constant above 1350 °C, meaning the complete melting and homogenization of the sample. This result along with DSC data gave a reference on controlling the temperature for the fiber spinning (Fig. 1B).

The morphology of MF was visualized by SEM. From the SEM images (Fig. 3A), we can see that the surface of all fibers is smooth and cylindrical. This observation originated from the melt shrinkage under the effect of surface tension during the fiber spinning. The diameter of the fiber decrease from 13.9 to 9.7 μm with increasing winding speed (Table 2), this is due to the fact that the melt undergoes a rapid lengthening and thinning process accompanied by solidification and external traction during the spinning.

The tensile strength of the fiber decreased from 1320 to 894 MPa with increasing winding speed, whereas the moduli of fibers were maintained at around 94 GPa. It should be mentioned here that when following the size-effect theory caused by the surface defects in a brittle material [25], the strength of the fiber should be increased with decreasing diameter. We speculated such deviation was caused by the microstructure of fiber under different winding speeds. The fiber is in an amorphous state and the atoms in the material are arranged in a long-range disordered way and stacked as dense as possible to form a random stacking microstructure [26]. Therefore, the denser microstructure results in an enhanced tensile strength of fiber. In addition, during the fiber-forming process, the lower winding speed resulted in a larger diameter of the fiber, and the cooling rate of the outside fiber was much higher than that in the inside. In this way, the atoms in the fiber have a longer time for relaxation, which leads to a higher tensile strength [27].

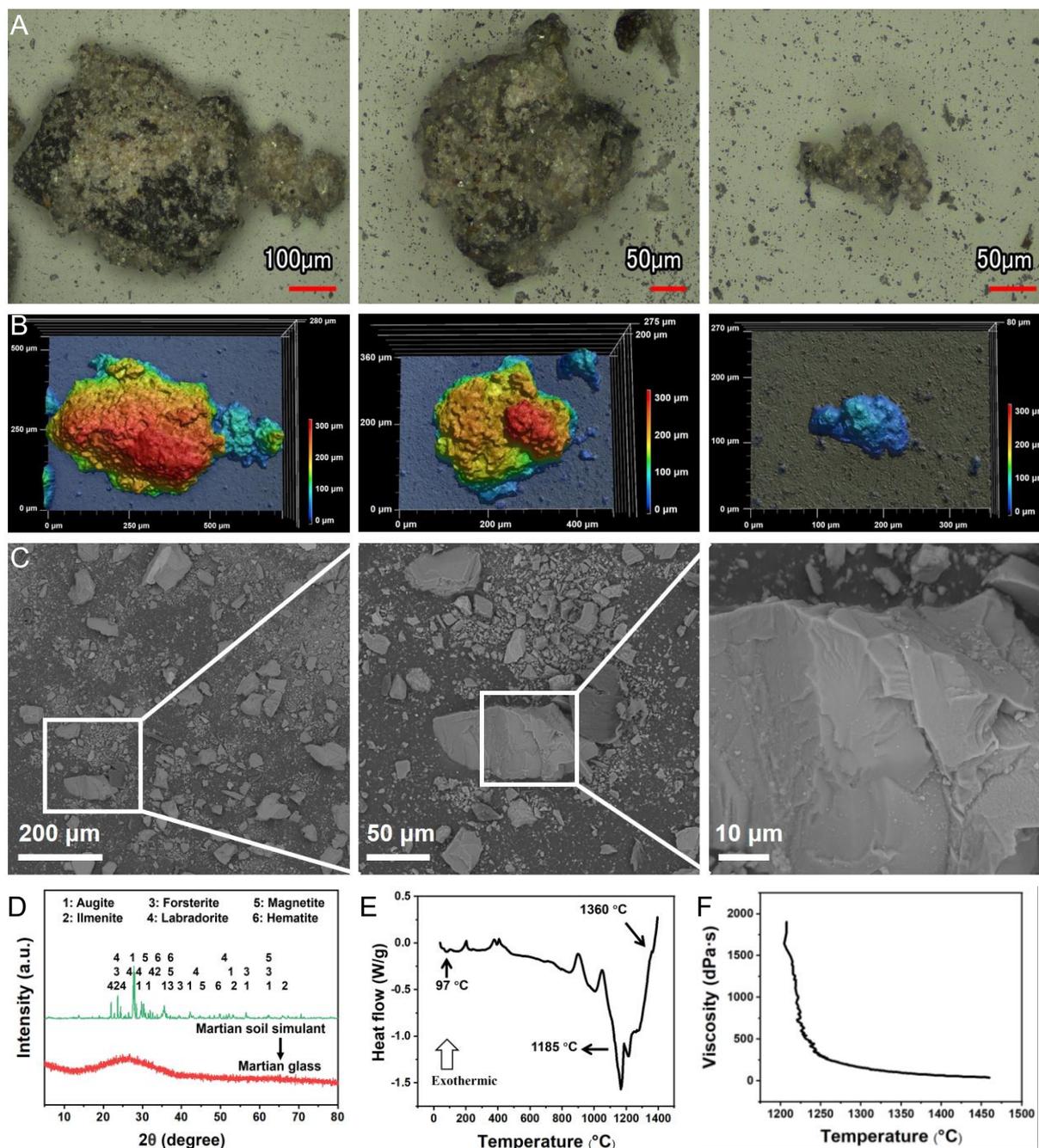

**Fig. 2.** Morphology and properties of Martian soil simulant (A and B: Typical particle morphology of Martian soil simulant with different particle sizes and magnifications; C: SEM images under different magnifications; D: Mineral phases of simulant and XRD spectra; E: DSC curve of the sample; F: Temperature-viscosity curve of the sample).

**Table 2** Tensile properties of MF produced at the different winding speeds.

| Sample | Winding speed (m/s) | Fiber diameter (μm) | Tensile strength (MPa) | Tensile modulus (GPa) | Elongation at break (%) |
|---|---|---|---|---|---|
| MF-2.5 | 2.5 | 13.9±1.4 | 1320±249 | 99±23 | 1.5±0.4 |
| MF-4.0 | 4.0 | 11.5±0.6 | 1081±247 | 94±20 | 1.3±0.3 |
| MF-6.0 | 6.0 | 9.7±0.8 | 990±317 | 94±15 | 1.1±0.3 |

To prove the above assumption, Raman spectroscopy was used to evaluate the fine structure of fiber samples. The broad band in the range of 800-1200 cm$^{-1}$ reflects the symmetric stretching vibration of Si-O tetrahedron, and this position was mainly composed of four spectral peaks caused by the symmetrical stretching vibration of bridging oxygen in terms of $Q^0$, $Q^1$, $Q^2$, and $Q^3$, corresponding to the Raman shifts at around 850-880 cm$^{-1}$, 900-920 cm$^{-1}$, 950-980 cm$^{-1}$ and 1050-1100 cm$^{-1}$, respectively [28]. The $Q^n$ is classified by the number of bridging oxygen in the Si-O tetrahedron, for example, $Q^3$ stands for silicon coordinated by three bridging oxygens, and $Q^0$ means the absence of bridging oxygen. The relative percentages of various structural units were analyzed by Gaussian fittings as shown in Fig. 3B, and calculated using the fitted peak areas as expressed in Equation 1 [29]:

$$Xn = \frac{An/Sn}{\sum_0^3 An/Sn} \quad (1)$$

Where $X_n$ is the relative percentages of various structural units $Q^n$, $A_n$ is the percentages of fitted peak areas of $Q^n$, and $S_n$ is the Raman scattering coefficient of $Q^n$ (Here $S_0$=1, $S_1$=0.514, $S_2$=0.242, $S_3$=0.09) [29].

The ratio of relative percentage of $Q^3$ to $Q^2$ ($X_3/X_2$) was used to describe the degree of polymerization of glass, and the larger the ratio is, the denser the glass structure presents [30]. By comparing the $X_3/X_2$ ratio of different samples, we can evaluate the packing compactness of microstructures in fiber samples. As shown in Table 3, the MF produced at 2.5 m/s has the highest $X_3/X_2$ ratio, and the ratio decreased with increasing winding speed, confirming that the fiber produced at a lower winding speed has a denser microstructure, this in turn leads to the higher tensile strength of the fiber.

**Table 3** Relative content of $Q^n$ of MF produced at the different winding speeds.

| Sample | $X_0$ | $X_1$ | $X_2$ | $X_3$ | $X_3/X_2$ | Average strength (MPa) |
|---|---|---|---|---|---|---|
| MF-2.5 | 0.09 | 13.61 | 60.32 | 25.98 | 0.43 | 1320 |
| MF-4.0 | 1.54 | 23.56 | 66.48 | 8.42 | 0.13 | 1081 |
| MF-6.0 | 1.77 | 22.79 | 70.85 | 6.61 | 0.09 | 990 |

The pair distribution function (PDF) data were collected from the neutron scattering source to further investigate the effect of winding speed on the microstructure of fiber. As shown in Fig. 3C, the first peak appeared in PDF for the two typical fibers (MF-2.5 and MF-6.0) is around 1.64 Å, which reflects the intra-tetrahedral Si-O distance. The peaks between 1.7-2.0 Å correspond to the distances of other glass network formers with oxygen, such as Al-O and Fe-O [31, 32]. As we can see from the curves, the value of PDF ($G_r$) of glass network former in MF-2.5 fiber are always higher than that produced at 6.0 m/s, meaning a tighter network formed by the tetrahedral structure in the sample. Additionally, we calculated the peak area (Inset in Fig. 3C) of the PDF, and found that this value for MF-2.5 (1.31) is 15% higher than MF-6.0 (1.11). Since the fibers produced at different winding speeds have the same chemical composition, we can determine the degree of network aggregation by directly comparing the peak areas, and a higher peak area represents a higher average $Q^n$ [33, 34], which is consistent with Raman results.

From the FT-IR curves of MF (Fig. 3D), we can see that all fiber samples show a pretty similar pattern. Specifically, the wide band centered at 3430 cm$^{-1}$ and the one at around 1630 cm$^{-1}$ correspond to the stretching vibration of hydroxyl group [35], and the strong band at around 1000 cm$^{-1}$ is ascribed to the anti-symmetric stretching vibration of Si-O-Si in the material [36]. The one at around 1270 cm$^{-1}$ indicates the stretching vibration of Al-O-Si network [35]. It should be mentioned here that the hydroxyl groups on fiber surface can adsorb and react with polymeric, ceramic or metallic matrices, giving the great potential as reinforcement to develop composites for the construction of Martian base.

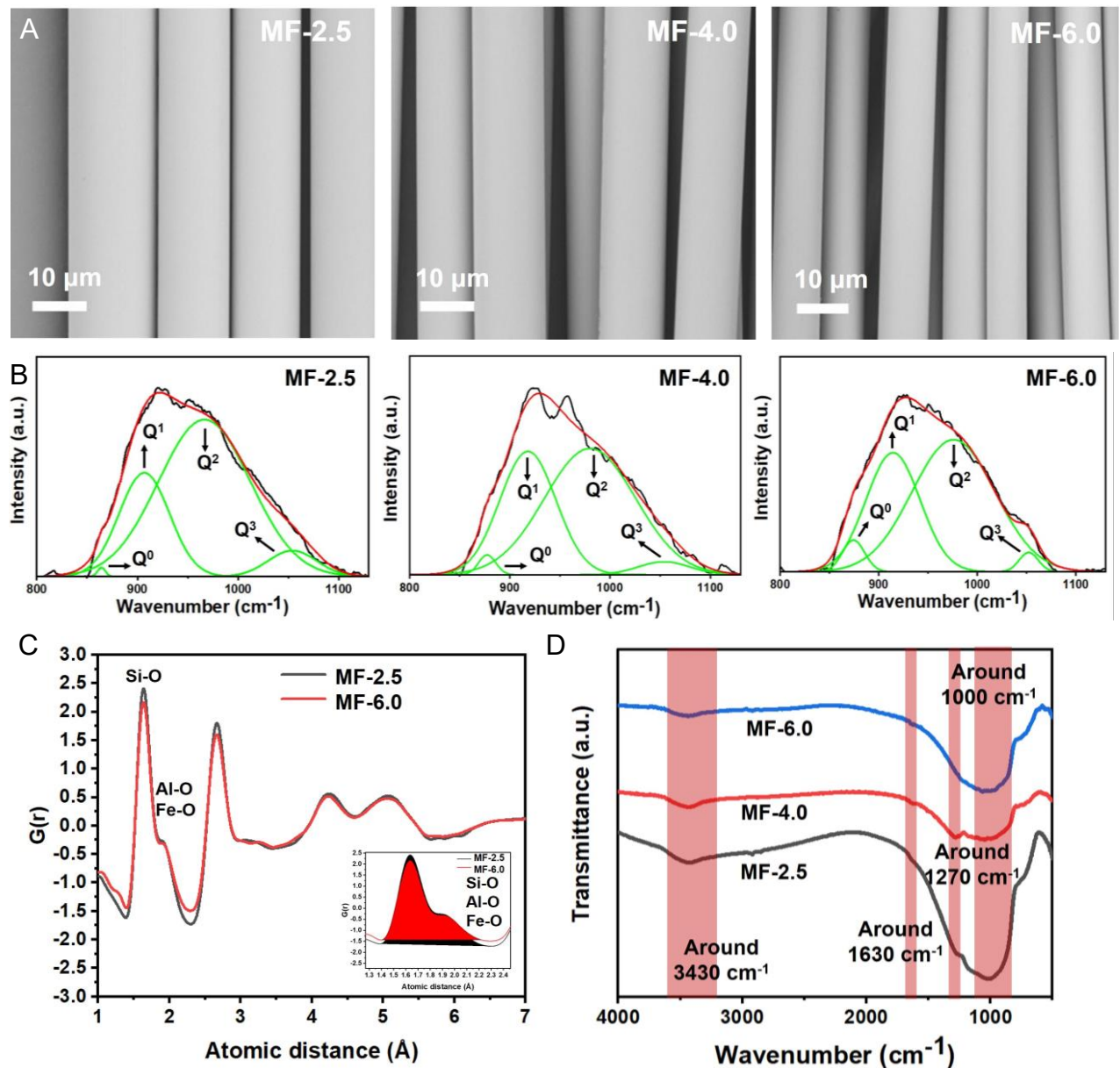

**Fig. 3.** Properties and fine structures of MF produced at the different winding speeds (A: Morphology; B: Raman spectra and fitting results, for all fitting results $R^2$>0.99; C: Partial distribution function; D: FT-IR spectra).

The environment of Mars is different from that on the Earth. Mars has a low gravity (1/3 $g$) and atmospheric pressure (0.7 KPa), as well as different atmospheric components (95.0% carbon dioxide, 3.0% nitrogen, 1.6% argon and 0.13% oxygen) [1, 37]. Such conditions will affect the fiber formation process. For the low gravity, previous studies showed that the glass prepared under a low $g$ was more homogeneous and resistant to the crystallization than that prepared on the Earth [38, 39]. The enhanced homogeneity of constituents can improve the uniformity of fiber and result in better mechanical performance. The better resistance to crystallization brings a positive effect on improving the continuity of fiber production and a more straightforward design for equipment. As for the atmosphere, the components of Mars are mainly inert gas (about 99.6%), which facilitates the maintenance of $Fe^{2+}$ in the Martian soil. This will facilitate the formation of fiber and its strength, as previous studies showed that the tensile strength of fiber increased with an increasing ratio of $Fe^{2+}/\Sigma Fe$ [40].

**Conclusions**
In summary, we demonstrated the feasibility of using Martian soil to get corresponding fiber. Based on the analysis on the chemical composition and properties of soil simulant, continuous fiber was produced under different winding speeds on a 50-hole fiber spinning facility. The fiber produced with a winding speed of 2.5 m/s shows the highest tensile strength of 1320 MPa, closing to the fiber prepared on the Earth used for the development of composite materials. The findings of current study confirm that continuous fibers can be obtained from Martian soil by melting and spinning, and the results offer a new way for the in-situ utilization of Martian resource to develop structural and functional materials for the construction of Martian base. There will be numerous challenges in achieving this goal, for example, the influence of Martian environment (lower gravity, inert atmospheric, etc) on the fiber properties, design and system integration of fiber-forming equipment. Definitely, there will be a bright future in this interesting research field with more participants and collaborations worldwide.

**Acknowledgements:**
This work was supported by Tianshan Talent Program for Scientific and Technological Innovation in Xinjiang 2022TSYCLJ0041 and Silk Road Innovation Fund under the framework of Frontier Technology Program of the 14th Five-Year Plan of CAS XTIPC-2023SLSZCXJJ-XX01.

**Compliance with ethics guidelines**
Ze-Shi Guo, Dan Xing, Xiong-Yu Xi, Cun-Guang Liang, Bin Hao, Xiaojia Zeng, Hong Tang, Huaican Chen, Wen Yin, Peng Zhang, Kefa Zhou, Qingbin Zheng and Peng-Cheng Ma declare that they have no conflict of interest or financial conflicts to disclose.